\documentclass{article}
\usepackage[inkscapeformat=png]{svg}
\usepackage{microtype}
\usepackage{graphicx}
\usepackage{subfigure}
\usepackage{booktabs} 

\usepackage{hyperref}

\usepackage[accepted]{icml2024}

\usepackage{amsmath}
\usepackage{amssymb}
\usepackage{mathtools}
\usepackage{amsthm}

\usepackage[capitalize,noabbrev]{cleveref}

\theoremstyle{plain}
\newtheorem{defn}{Definition}[section]

\theoremstyle{definition}

\theoremstyle{remark}

\usepackage[textsize=tiny]{todonotes}

\icmltitlerunning{WhaleNet: Wavelet High Adaptive Learning Ensemble Network}

\begin{document}

\twocolumn[
\icmltitle{WhaleNet: a Novel Deep Learning Architecture for Marine Mammals Vocalizations on Watkins Marine Mammal Sound Database}



\icmlsetsymbol{equal}{*}

\begin{icmlauthorlist}

\icmlauthor{Alessandro Licciardi}{yyy,comp}
\icmlauthor{Davide Carbone}{yyy,comp}

\end{icmlauthorlist}

\icmlaffiliation{yyy}{Department of Mathematical Sciences, Politecnico di Torino, Torino, Italy}
\icmlaffiliation{comp}{Istituto Nazionale di Fisica Nucleare (INFN), Sezione di Torino, Torino, Italy}

\icmlcorrespondingauthor{Alessandro Licciardi}{alessandro.licciardi@polito.it}

\icmlkeywords{Machine Learning, ICML}

\vskip 0.3in]



\printAffiliationsAndNotice{}  

\begin{abstract}

Marine mammal communication is a complex field, hindered by the diversity of vocalizations and environmental factors. The Watkins Marine Mammal Sound Database (WMMD) constitutes a comprehensive labeled dataset employed in machine learning applications. Nevertheless, the methodologies for data preparation, preprocessing, and classification documented in the literature exhibit considerable variability and are typically not applied to the dataset in its entirety. This study initially undertakes a concise review of the state-of-the-art benchmarks pertaining to the dataset, with a particular focus on clarifying data preparation and preprocessing techniques. Subsequently, we explore the utilization of the Wavelet Scattering Transform (WST) and Mel spectrogram as preprocessing mechanisms for feature extraction. In this paper, we introduce \textbf{WhaleNet} (Wavelet Highly Adaptive Learning Ensemble Network), a sophisticated deep ensemble architecture for the classification of marine mammal vocalizations, leveraging both WST and Mel spectrogram for enhanced feature discrimination. By integrating the insights derived from WST and Mel representations, we achieved an improvement in classification accuracy by $8-10\%$ over existing architectures, corresponding to a classification accuracy of $97.61\%$.
\end{abstract}

\section{Introduction}

Marine mammals, which include species such as whales, dolphins, and seals, are celebrated for their intricate communication systems, crucial for survival and social interactions. Despite the significance of these communication systems, understanding them remains challenging due to the diverse range of vocalizations, behaviors, and environmental factors involved \cite{watkins1985sensory}\cite{dudzinski2009communication}. Recent research efforts have increasingly turned towards the use of machine learning (ML) to analyze and decipher communication patterns between marine mammals \cite{mazhar2007vocalization} \cite{bermant2019deep}. The application of AI and ML enables researchers to classify vocalizations effectively, monitor movements, and gain insights into behavior and social structures \cite{mustill2022speak}. In addition, these technologies support ecological studies by correlating whale vocalizations with environmental factors, providing valuable information on behavioral patterns and social structures. Real-time monitoring establishes early warning systems for conservation efforts, helping mitigate the impact of human activities on whale populations \cite{croll2001effect}\cite{gibb2019emerging}.\\
A significant resource in the study of marine mammal communication is the Watkins Marine Mammal Sound Database (WMMD) \cite{sayigh2016watkins}. Spanning seven decades, this collection of recordings encompasses various species of marine mammals and holds immense historical and scientific value. Although the WMMD serves as a renowned reference dataset for studying vocalizations, it presents challenges for classification, including variability and complexity in vocalizations, environmental noise, and data scarcity for certain species.\\
Current state-of-the-art benchmarks heavily rely on deep learning \cite{ghani2023global} or peculiar data preparation and preprocessing \cite{murphy2022residual}\cite{hagiwara2023beans}\cite{hagiwara2023aves}. Moreover, most of current works usually tackle just portion of the full dataset, as for instance very few classes \cite{lu2021detection} or the "best of" subset \cite{hagiwara2023beans}. Moreover, the main preprocessing methods are based on Short Time Fourier Transform \cite{roberts1987digital_signal_processing} and other specifications. Addressing these issues, we introduce the Wavelet Scattering Transform (WST) \cite{mallat2012group}\cite{bruna2013invariant} in our work. Regarded as the mathematical counterpart of convolutional layers in deep networks, WST boasts invariance and stability properties concerning signal translation and deformation, qualities absent in standard preprocessing. Furthermore, the structure of the scattering coefficients proves valuable in providing a physical interpretation of multiscale processes, especially in the context of complex natural sounds \cite{khatami2018origins}.\\
The significance of the data set extends beyond biology, representing a notable example of natural time series. The preprocessing and statistics of such objects present a long-standing challenge in data science, from the early methods based on Fourier analysis to modern AI-based tools \cite{fu2011review}\cite{aghabozorgi2015time}. WST has found application in various physical datasets, contributing to advances in understanding multiscale and multifrequency processes that are challenging to address with standard Fourier techniques \cite{bruna2019multiscale}\cite{cheng2020new}\cite{glinsky2020quantification}.\\
In this study we focus on WMMD and:
\begin{itemize}
    \item we collect a review of data preparation, preprocessing and classification methods used in literature which can be potentially important for bioacustics community;
    \item we provide a novel detailed and public pipeline for data preparation of WMMD, highlighting the possibility of using WST as alternative preprocessing method;
    \item we propose WhaleNet, a novel deep architecture with residual layers that ensembles WST and Mel spectrogram, demonstrating higher classification accuracy compared to existing benchmarks.
\end{itemize}
In Table \ref{tab:results} we report a short summary of the accuracy results for the classification task, as opposed to the existing benchmarks. The code for the present work is available on the public GitHub repository \href{https://github.com/alelicciardi99/whalenet/tree/main}{\texttt{whalenet\_vocalization\_classification}}.
\begin{figure*}[h]
    \centering   \includegraphics[width=0.8\textwidth]{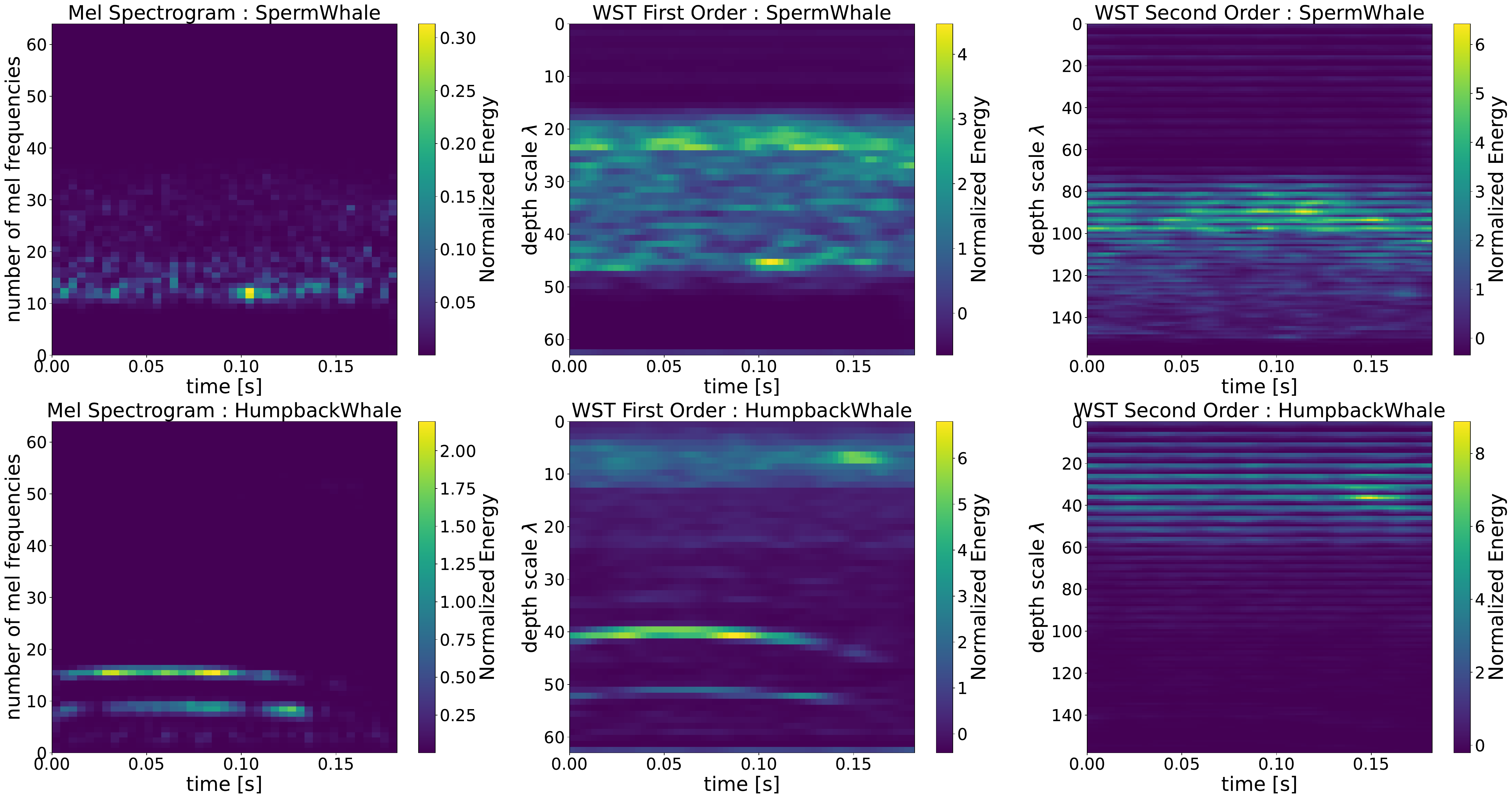}
    \caption{\textit{From left:} Mel spectrogram, WST of first and second order for vocalizations of two different species of whales. The displayed WSTs correspond to the choice $(J,Q)=(7,10)$. Focusing on the second row, it is graphically evident the correspondence of a high-depth scale for WST with low frequency in the spectrogram. Mel spectrogram appears to be more coarse-grained with respect to first-order WST, even if the overall heatmaps appear to be similar. Each figure is resized to be squared for visualization purposes. The shapes of the images in each row are, from left, respectively 41$\times$64 for Mel spectrogram and 53$\times$63 and 158$\times$63 for first and second order WST.}
    \label{fig:wst_mel}
\end{figure*}
\section{Preprocessing techniques }
\subsection{STFT and Mel Spectrogram}
Spectrogram representation is one of the most common technique used in 1D signal representation theory, cfr. \cite{roberts1987digital_signal_processing}. It provides information on the energy spectrum in the time-frequency domain $(t,\omega)$ and is based on the \textit{Short Time Fourier Transform} (STFT).
Let us briefly recall the definition of STFT: we suppose that the time variable $t$ is a positive real number, i.e. $t\in[0,+\infty)$. Let us fix a function $h(t)$ called \textit{window function}, most common choices being \textit{Hann window} or \textit{Gaussian window}. Hann window, with support length $T>0$, has the following form
\begin{equation}
\label{eq:hann}
    h(t)=a\cos^2\left(\frac{\pi t}{T}\right)\mathbf{1}_{\{|t|\leq T/2\}}(t)\,
\end{equation}
while Gaussian window is a centered Gaussian function with amplitude $a$ and spread $\sigma$, i.e.
\begin{equation}
\label{eq:gauss}
    h(t)=a\exp\left(-\frac{t^2}{2\sigma^2}\right)\,.
\end{equation}
As one can infer from the name, a window function is usually chosen to be localized in time domain, and can also be compactly supported as \eqref{eq:hann}. We can then recall the following definition:
\begin{defn}\label{STFT}
    For a given signal $x(t)$ and a fixed window function $h(t)$, the \textbf{Short Time Fourier Transform} is defined as
    \begin{equation}
        \mathbf{STFT}\{x\}(t,\omega)= \int _{-\infty}^\infty x(\tau)h(\tau-t)e^{-i\omega\tau} \,d\tau.
    \end{equation}
\end{defn}

Note that STFT is strictly related to the Fourier transform operator $\mathcal{F}$, due to the immediate relation \begin{equation}
    \mathbf{STFT}\{x\}(t,\omega)=\mathcal{F}\{x(\tau)h(\tau-t)\}(\omega)
\end{equation}
i.e. the Fourier transform of the signal $x(\tau)$ multiplied by a moving window $h(\tau-t)$, for any $t>0$.
A trivial extension of the definition to the discrete time case is possible, by replacing the integral with an infinite summation. Given the STFT we recall the definition of spectrogram
\begin{defn}\label{def_spectrogram}
For any $t>0$ and $\omega>0$, and for a chosen window $h(t)$ the \textbf{spectrogram} of a signal $x$ is defined as the power spectrum of $x(\tau)h(\tau-t)$, i.e.
\begin{equation}
|X(t,\omega)|^2=|\mathbf{STFT}\{x\}(t,\omega)|^2\,,
\end{equation}
\end{defn}

The Mel spectrogram \cite{rabiner2010theory}, often employed in audio signal processing, involves a transformation of the spectrogram introduced in Definition \eqref{def_spectrogram} to a Mel frequency scale. This scale is designed to mimic the human ear's nonlinear frequency perception.
For a given signal $x(t)$ and a chosen window function $h(t)$, the Mel spectrogram is defined as the power spectrum of the signal transformed to the Mel frequency scale. It provides a detailed representation of the signal's energy distribution across both time and Mel frequency variables. The first step in computing the Mel spectrogram involves defining a set of triangular filters, often referred to as the Mel filter bank. These filters are spaced along the Mel frequency scale and overlap to capture the nonuniform nature of human hearing. This scaling choice is highly motivated for natural sounds and has been used for preprocessing since the first application to classification of labeled sounds \cite{lee2006automatic}. Informally, an analysis of the signal that is based of an ear-like preprocessing should simplify classification. Let $N$ be the number of filters in the Mel filter bank and $f(m)$ be the center frequency of the $m$-th filter. The Mel frequency $m$ corresponding to a given frequency $\omega$ is computed using the formula:

\begin{equation}
m = 2595 \cdot \log_{10}\left(1 + \frac{\omega}{700}\right).
\end{equation}

The center frequency $f(m)$ in Hertz corresponding to a Mel frequency $m$ is then given by:

\begin{equation}
f(m) = 700 \cdot (10^{m/2595} - 1).
\end{equation}

Each triangular filter $H_m(\omega)$ is defined as
\begin{equation}
H_m(\omega) =
\begin{cases}
0 & \text{if } \omega < f(m-1) \\
\frac{\omega - f(m-1)}{f(m) - f(m-1)} & \text{if } f(m-1) \leq \omega \leq f(m) \\
1 - \frac{\omega - f(m)}{f(m+1) - f(m)} & \text{if } f(m) \leq \omega \leq f(m+1) \\
0 & \text{if } \omega > f(m+1)
\end{cases}
\end{equation}
The Mel spectrogram is computed by summing the energy in each triangular filter bank applied to the magnitude of the Short Time Fourier Transform (STFT) of the signal:

\begin{equation}
\text{Mel Spectrogram}(t, m) = \sum_{k=0}^{N-1} |X(t, \omega_k)|^2 \cdot H_m(\omega_k),
\end{equation}

where $N$ is the number of frequency bins in the STFT, $X(t, \omega_k)$ is the STFT magnitude at time $t$ and frequency bin $\omega_k$, and $H_m(\omega_k)$ is the value of the $m$-th Mel filter at frequency bin $\omega_k$.
\subsection{Wavelet Scattering Transform}
\begin{figure}
    \centering
    \includegraphics[width=1\linewidth]{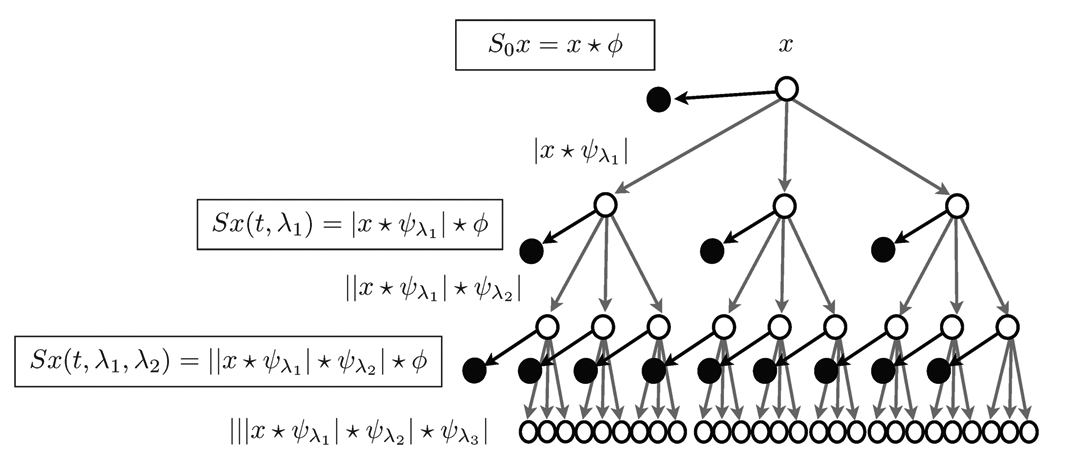}
    \caption{Wavelet Scattering Transform as an iterative process; image taken from \cite{anden2014deep}. In their notation the signal is $x(t)=h(t)$ the path $p$ at depth $m$ is explicited in parentheses as a tuple $(\lambda_1,\dots,\lambda_m)$. Each black dot corresponds to a scattering coefficient.}
    \label{fig:wst}
\end{figure}

The Wavelet Scattering Transform (WST) \cite{mallat2012group} stands as a mathematical operator capable of yielding a stable and invariant representation for a given signal. Specifically, when certain conditions are met \cite{bruna2013invariant}, the resulting representation exhibits translation invariance, resistance to additive noise (i.e., it remains non-expansive), and stability to deformations. The latter property is formally expressed as Lipschitz continuity under the influence of $C^2$-diffeomorphisms in its original derivation. Integrating a representation operator with these advantageous characteristics into a machine learning framework has the potential to significantly reduce the computational burden involved in training classification algorithms \cite{brunaphd}. Since its derivation has been proposed very recently, in this section we provide an extended summary of definition and properties of WST for 1D signals (n.b. an extension to higher dimensions can be found, for instance, in \cite{bruna2013invariant}). \\
Let $\psi \in L^2(\mathbb{R}, dx)$ be a function, called \textit{mother wavelet}, for a fixed scale factor $a>1$ and for any $j\in \mathbb{Z}$, the $j-$th wavelet is defined as
\begin{equation}\label{defwst}
    \psi_{a^j}(t)=a^{- j}\psi(a^{-j} t)
\end{equation}
Let $\lambda= a^j $ be the scaling-rotation operator, \eqref{defwst} can be redefined in terms of $\lambda$ as
\begin{equation}
    \psi_\lambda(t)= \lambda^{-1} \psi(\lambda^{-1} t)\,.
\end{equation}
To build an intuitive connection with STFT, $a$ is analogous to the width used for Hann or Gaussian windows. Concerning the choice of the mother wavelet, we refer in the following to the Morlet wavelet \cite{mallat1999wavelet}. In practice, in the usual definition of WST, they define $Q\in\mathbb{N}$ such that $a=2^{1/Q}$; this will play a role of a hyperparameter.\\
In order to construct the wavelet scattering operator, we fix the depth $J\in\mathbb{N}$ and let $\Lambda_J=\{\lambda=a^j  \,: |\lambda|=a^j\leq2^J\} $ be the set of scattering indexes. 
Then, we introduce a scaled low-pass filter $\phi_J(t) = 2^{-J}\phi(2^{-J}t)$, where $\phi(t)$ is a Gaussian $\mathcal{N}(0,\sigma^2)$ with $\sigma=0.7$,
and a path $p=(\lambda_1,\dots.\lambda_m)$, $\lambda_i \in \Lambda_J$ which is any tuple of length $m$ build using the scattering indexes; the wavelet scattering coefficient along a path $p$ is defined as 
\begin{equation}
S_J[p]x(u)=U[p]x\star\phi_J(t) = \int_{-\infty}^\infty U[p]x(\tau) \phi_J(t- \tau) d\tau\,,
\end{equation}
where
\begin{equation}
    U[p]x=U[\lambda_m]\dots U[\lambda_1]x=|\dots|x\star\psi_{\lambda_1}|\star\psi_{\lambda_2}|\dots|\star\psi_{\lambda_m}| \,.
\end{equation}
For the conducted experiments we couple \textit{Morlet wavelets} with a Gaussian low-pass filter \cite{mallat1999wavelet}.\\
Let us clarify the definition of WST in layman's terms: by a simple combinatorial argument, the longer is the path, the larger is the number of combinations of scattering indexes, and more precisely, one has the characteristic tree structure, as one can see in Figure \ref{fig:wst}. Each black dot corresponds to a scattering coefficient and is usually referred to the coefficient for fixed $m$ as the $m$-\textit{order} scattering coefficients. In analogy to the spectrogram representation, it is usual to plot the coefficients of the same order on a single heatmap, having time and $j$ on the axis, see Figure \ref{fig:wst_mel}. Notice how $J$ is another free hyperparameter whose effect is to increase the cardinality of $\Lambda_J$, hence the number of coefficients per order. To practically infer the importance of the order, following \cite{mallat2012group} we introduce the path set up to length $m$, $\Lambda_J^m=\{(\lambda_1,\dots,\lambda_m):\,|\lambda_i|=a^j\leq 2^J\}$, it is possible to define the induced norm of the scattering operator over the set $\mathcal{P}_J=\bigcup \Lambda_J^m$, i.e.
\begin{equation}\label{scattering_norm}
    \|S_J[\mathcal{P}_J]x\|=\sum_{p\in \mathcal{P}_J}\|S_J[p]x\|
\end{equation}
where $\|\cdot\|$ stands for the $L^2-$norm. For fixed $J$ and $Q$, and given the definition of $\Lambda_J^m$.
One could be concerned about the depth requested in practice, but in different experiment \cite{brunaphd} it has been showed that just $2$ or $3$ orders, also referred to as layers, of WST are sufficient to represent around $98\%$ of the energy of the signal. Indeed the energy of each layer, i.e. $\|U[\Lambda_J^m]\|$, is empirically observed to rapidly converge to zero. Thus, usually no more than two orders need to be computed to capture most of the information contained in the signal.
\section{Training and test datasets set-up}
In this section, our objective is to conduct a comprehensive comparison of the data analysis between the Mel spectrogram and the WST. We emphasize that the pipeline for this comparison is entirely general and could potentially be extended to any temporal series. Notably, the application of WST as a theoretical tool is already prevalent in diverse fields such as cosmology \cite{valogiannis2022towards} and field theory \cite{marchand2022wavelet}. As a widely acknowledged principle in the literature \cite{bruna2013invariant}, WST is preferable to STFT methods when their performances are comparable, primarily due to the invariance properties that facilitate cross-signal interpretation.  

\subsection{Watkins Marine Mammal Sound Database}
In this study, we use the expansive Watkins Marine Mammal Sound Database \cite{sayigh2016watkins} as a foundational dataset for our research. The database, spanning from the 1940s to the 2000s, offers a rich collection of over 2000 recordings that encompass more than 60 species of marine mammals, serving as a valuable resource for marine mammal detection in Passive Acoustic Monitoring (PAM) data. Specifically, three directories are available within the database (website: \href{https://cis.whoi.edu/science/B/whalesounds/index.cfm}{https://cis.whoi.edu/science/B/whalesounds/index.cfm}): {\it 'Best of' Cuts}, {\it All Cuts}, and {\it Master tapes}, containing recordings of varying quality and length.
\begin{figure}[h]
    \centering
    \includegraphics[width = \linewidth]{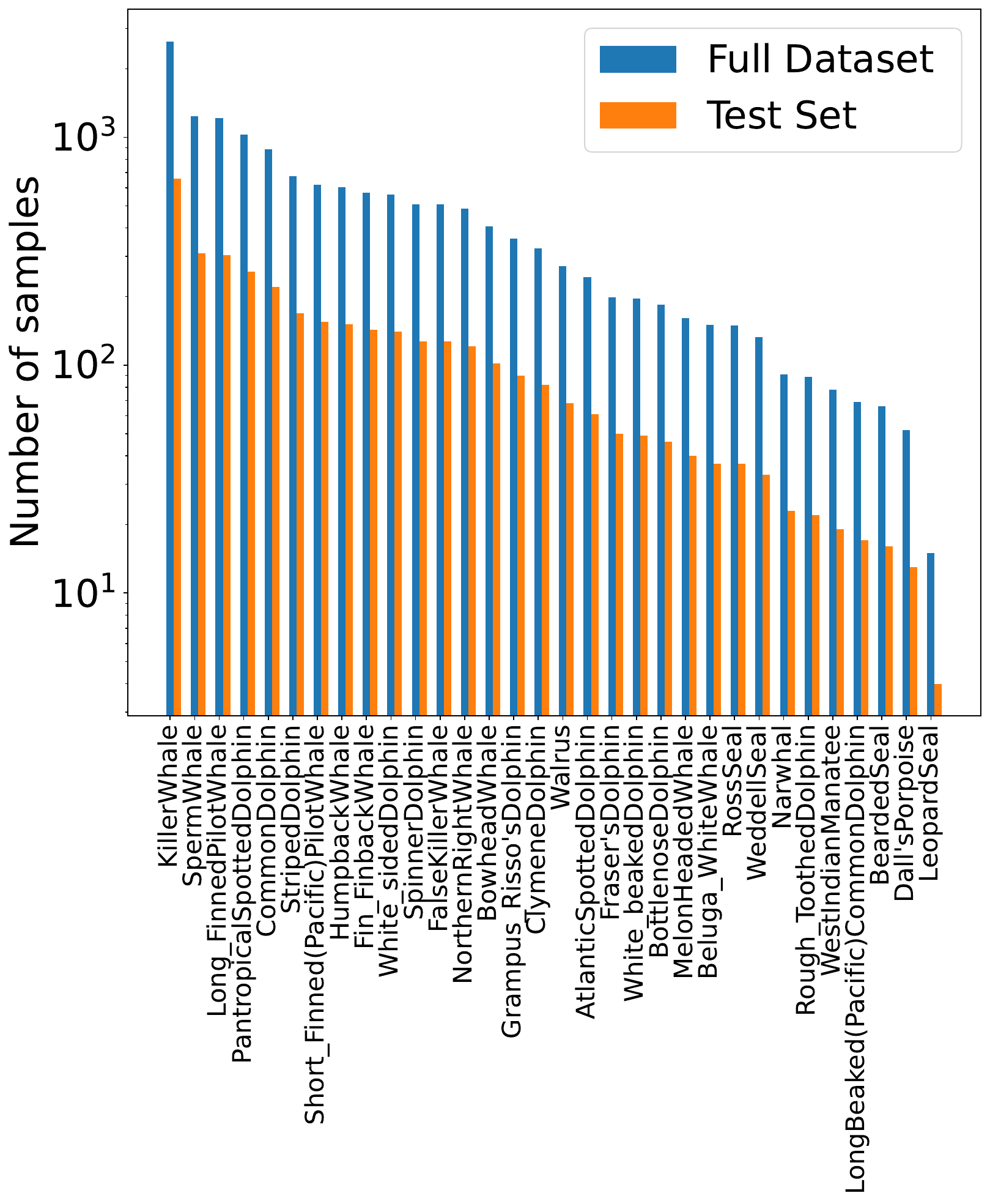}
    \caption{Number of samples per class after data preparation and elimination of duplicates, in log-scale and sorted in decreasing order. The dataset is very imbalanced: the most represented class contains $2637$ instances, while the smallest one just $15$.}
    \label{fig:distri}
\end{figure}
\floatname{algorithm}{\color{black}Step}
\begin{algorithm}[h]
\caption{\color{black}Align the signal with padding or cutting}\label{alg:cutter}
\begin{algorithmic}
    \STATE \textbf{Input:} signal $x \in \mathbb{R}^K$, output signal length $T$
    \vspace{1em}
    \IF{$T \geq K$ } 
    \STATE $t_c \gets \lfloor K/2\rfloor$
    \STATE $t_l \gets T - t_c$
    \STATE $t_r \gets T - t_l$
    \STATE $x' \gets x[t_c-t_l: t_c + t_r]$ \COMMENT{cutting the original signal around its central time}
    \ELSE  
    \STATE $\Delta \gets K - T$
    \STATE $ t_l \gets \lfloor \Delta/2\rfloor$
    \STATE $t_r \gets \Delta - t_l$
    \STATE $ x' \gets (\mathbf{0}_{t_l}, x, \mathbf{0}_{t_r})$ \COMMENT{center the original signal and then add zeros on both sides}
    \ENDIF
    \vspace{1em}
    \STATE \textbf{Output:} transformed signal $x' \in \mathbb{R}^T$
\end{algorithmic}
\end{algorithm}

\begin{algorithm}[h]
\caption{\color{black}Standardize Signal}\label{alg:std}
\color{black}
\begin{algorithmic} 
\STATE \textbf{Input:} original signal $x \in \mathbb{R}^T$
\vspace{1em}
\STATE $\hat{\mu} \gets \dfrac{\sum_{t = 1}^T x(t)}{T} $\COMMENT{Compute the sample mean}
\vspace{.5em}
\STATE $\hat{\sigma}^2 \gets \dfrac{\sum_{t = 1}^T (x(t)-\hat{\mu})^2}{T-1}$ \COMMENT{Compute the sample variance}
\vspace{.5em}
\STATE $x'(t) \gets \dfrac{x(t) - \hat{\mu}}{\hat{\sigma}} \,\,\,\, t = 1,\dots, T$ \COMMENT{Standardization}
\vspace{1em}
\STATE \textbf{Output:} standardized signal $x' \in \mathbb{R}^T$
\end{algorithmic}
\end{algorithm}
\floatname{algorithm}{\color{black}Algorithm}
\begin{algorithm}[h]
\caption{\color{black}Data Preparation and Preprocessing}\label{alg:preprocessing}
\color{black}
\begin{algorithmic}
\STATE\textbf{Input:} original signal $x_i \in \mathbb{R}^K$, target signal length $T$, representation operator $\Phi$, i.e. WST or Mel spectrogram
\vspace{1em}
\STATE $x_i \gets \text{Align}(x_i,T)$
\COMMENT{use Algorithm \ref{alg:cutter} to center and cut, or pad, the signal up to length $T$}
\vspace{.5em}
\STATE $x_i \gets \text{Standardize}(x_i)$
\COMMENT{use Algorithm \ref{alg:std} to standardize the signal}
\vspace{.5em}
\STATE $\phi_i \gets \Phi[x_i] \in \mathbb{R}^\Theta$ \COMMENT{compute WST or Mel spectrogram}
\vspace{1em}
\STATE\textbf{Output:} transformed signal $\phi_i \in \mathbb{R}^\Theta$
\end{algorithmic}
\end{algorithm}
In the present work, we consider the {\it All Cuts} part that comprises 15,554 samples collected over 70 years by the Woods Hole Oceanographic Institution, representing sounds produced by 51 marine mammal species. This choice is different from most of the benchmarks present in literature on classification task, where {\it 'Best of' Cuts} part is commonly used (cfr. \cite{lu2021detection,hagiwara2023beans,murphy2022residual,ghani2023global}).
\subsection{Data processing}
Challenges in the dataset include data heterogeneity due to different sensors and class-wise imbalance, leading us to follow the approach in \cite{bach2023classifying} by excluding classes with fewer than 50 samples, reducing the species to 32, as depicted in Figure \ref{fig:distri}. A detailed examination revealed over 300 repeated samples, some with different labels. Consequently, we removed duplicate signals, resulting in 14,767 unique signals. Another necessary preprocessing step concerns the sample rate. This quantity varies from a minimum of 320 Hz to a maximum of 192 kHz across the dataset.\\
In order to tackle this dishomogeneity, we follow the approach of \cite{lu2021detection}, which consists of resampling every signal at a fixed frequency. They chose 10 kHz but, for such a choice, 89.4\% of recordings would need to be down-sampled. To avoid an excessive loss of information for data points recorded with high sample rate, we select instead the median of the sample rates in the dataset, that is 47.6 kHz, as fixed frequency. \\
\textbf{Step 1:} to address varying signal lengths, we aligned and centered the time series, fixing the number of time stamps at 8,000. Signals longer than 8,000 retained central points, while shorter ones were padded with equal zeros on both sides. This length is significantly shorter with respect to related works in mammal vocalizations \cite{murphy2022residual,ghani2023global}, yielding less memory and computational overload for storing the signals and for spectrogram computation.\\
\textbf{Step 2:} Each signal was standardized, ensuring zero sample mean and unitary variance. Since the measurements are performed with different instruments, this step is important towards a uniformization of the dataset.\\
\textbf{Step 3:} For each signal, the Wavelet Scattering Transform (WST) up to the second order, and the Mel spectrogram, were computed. This study explores various combinations of WST hyperparameters, specifically the depth scale parameter $J$ and the resolution $Q$ (where $(J,Q)\in \{(7,10), (6,16)\}$), to capture diverse signal characteristics. The configuration $(6,16)$ is particularly adapted to the human auditory frequency range, as utilized in Free Spoken Digits classification \cite{andreux2020kymatio}. The zeroth-order WST, which provides no informative content, was excluded from the analysis. As far as the dimensions of the resultant images is concerned, for the configuration $(J,Q)=(7,10)$, the dimensions for the first and second order are 53$ \times$63 and 158$ \times$63, respectively; for the configuration $(J,Q)=(6,16)$, the dimensions for the first and second order are 63$ \times$125 and 158$ \times$125, respectively. Each order was normalized to the median, adhering to a standard procedure employed in other contexts for spectrograms, as referenced in \cite{gwpy}. Regarding the Mel spectrogram, the number of Mel frequencies was fixed at 64 to mitigate undesired border effects. The \textit{hop length} parameter was set to 200, resulting in a single-channel Mel spectrogram dimension of 41$ \times$64 for each signal. Analogous to the WST methodology, each spectrogram was normalized. In Figure \ref{fig:wst_mel}, we present examples of Mel spectrogram and WST of first and second order obtained following the described data preparation pipeline. 
\subsection{Training and Test Datasets}
To ensure a rigorous validation experiment, the cleaned and preprocessed dataset was split into two distinct subsets. Specifically, 75\% of the data samples were allocated for model's training, while the remaining 25\% were reserved for validation. Given the pronounced imbalance within the dataset, stratification was employed in the preparation of both the test and training sets. As per standard practice, generalization capability is assessed using the validation set, which is strictly excluded from the backpropagation process.
\subsection{Software and Computational Resources}
The Mel spectrogram is computed utilizing the \texttt{Torchaudio} Python library, whereas the Wavelet Scattering Transform (WST) is performed using the \texttt{Kymatio} Python library \cite{andreux2020kymatio}. The training of the neural networks is executed on GPUs, specifically the \texttt{RTX8000 NVIDIA}, available through the high-performance computing (HPC) facilities at New York University and Politecnico di Torino.
\section{Model Architecture Design}
In this section, we provide a detailed description of the classification algorithm, including both the architectures and the training setup used.
\subsection{Residual Learning}
\begin{figure}[h]
    \centering
    \includegraphics[width = \linewidth]{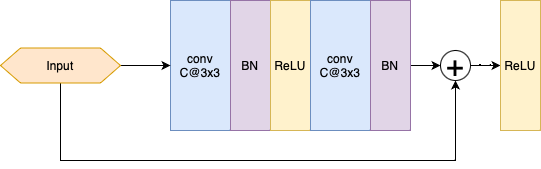}
    \caption{Structure of the residual block used in the full architecture \ref{fig:archi}. The acronym "BN" stands for batch normalization.}
    \label{fig:block}
\end{figure}

\begin{figure}[h]
    \centering
    \includegraphics[width = \linewidth]{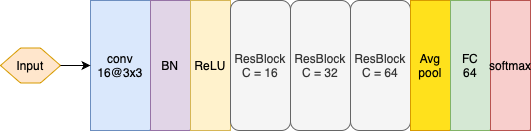}
    \caption{Structure of the architecture employed in the classification task. The residual blocks are unfolded in Figure \ref{fig:block}. The acronym "BN" stands for batch normalization, while "FC 64" denotes a fully connected layer with input dimension $64$ and output dimension $32$, corresponding to the number of classes. The number of trainable parameters is $176400$. For a comparison, AlexNet \cite{krizhevsky2017imagenet}, which is employed in \cite{lu2021detection} for a different classification task on WMMD, has 62.3 million of parameters.}
    \label{fig:archi}
\end{figure}

The ResNet architecture \cite{he2016deep} is widely used in deep learning applications due to its ability to train very deep networks effectively. The key innovation of ResNet is the use of residual blocks, which are defined as follows:
\begin{equation}
\mathbf{y} = \mathcal{F}(\mathbf{x}; W) + \mathbf{x}
\end{equation}
where \(\mathbf{x}\) and \(\mathbf{y}\) are the input and output of the block, and \(\mathcal{F}\) represents the residual mapping to be learned, and $W$ the model parameters. The addition of the input \(\mathbf{x}\) to the output of \(\mathcal{F}\) helps to mitigate the vanishing gradient problem and enables the training of deeper networks. This architecture is particularly useful in various deep learning tasks such as image classification, object detection, and semantic segmentation, where the depth of the network can significantly impact performance. By allowing gradients to flow through the network more effectively, ResNet facilitates the training of networks with hundreds or even thousands of layers, thereby improving accuracy and robustness in complex tasks. For our task, we propose a deep residual architecture composed of fundamental modules, as illustrated in Figure \ref{fig:block}. Specifically, each block consists of two convolutional layers, interspersed with Batch Normalization \cite{ioffe2015batch} and ReLU activation functions \cite{nair2010rectified}. Prior to the addition of the residual, an additional Batch Normalization layer is applied, followed by a final activation layer. We designed an ad-hoc convolutional architecture utilizing residual blocks, as summarized in Figure \ref{fig:archi}, capable of handling inputs of varying sizes without altering the total number of parameters. The initial layer is a $16 \times 3 \times 3$ convolutional layer, followed by batch normalization and activation layers. Further feature extraction is achieved through three residual blocks with increasing numbers of channels (16, 32, and 64). Finally, each channel is averaged, resulting in a flattened 54-dimensional feature vector, which is then processed by a fully connected neural network.

\subsection{WhaleNet Architecture}
\begin{figure}[h]
    \centering
    \includegraphics[width = .6\linewidth]{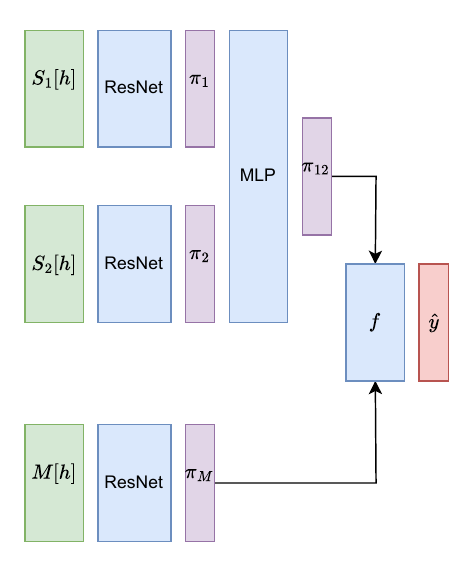}
    \caption{WhaleNet architecture. For a given input signal $h$, separately order 1 WST $S_1[h]$, order 2 WST $S_2[h]$ and Mel spectrogram $M[h]$ are fed to the ResNet model. Then output probabilities $\pi_1$ and $\pi_2$ are merged with a multi-layer perceptron, obtaining the WST merged probability output $\pi_{12}$. Then we propose different merging methods, namely $f$ to obtain the final prediction -- in particular element-wise maximum, hard convex combination $\lambda \pi_{12} + (1-\lambda) \pi_M$, or a fully connected MLP}
    \label{fig:fullnet}
\end{figure}

To fully exploit the feature extraction capabilities of both the WST and the Mel spectrogram, we propose a sophisticated architecture, called \textbf{WhaleNet}, that processes these representations separately, as illustrated in Figure \ref{fig:fullnet}. Using three ResNets in parallel, we extract three probability prediction vectors: $\pi_1$, $\pi_2$, and $\pi_M$. Following the principles of ensemble learning \cite{dong2020survey}, the WST prediction $\pi_{12}$ is obtained by training a multilayer perceptron (MLP) on the concatenated probability vector $[\pi_1, \pi_2]$.
To derive the prediction of the final class, we combine the information from the WST domain ($\pi_{12}$) and the Mel domain ($\pi_M$). We explore three methods to merge this information: \textit{max}, \textit{hard merge}, and \textit{MLP merge}.
The \textit{max} merge method involves taking the predicted class as the $\arg \max$ of the two stacked vectors. The \textit{hard merge} method \cite{bahaadini2018machine} computes the final probability as a convex combination of the two vectors, $\lambda \pi_M + (1 - \lambda)\pi_{12}$, where the optimal $\lambda$ is determined by grid search. Lastly, the \textit{MLP merge} method \cite{bahaadini2018machine} involves training a small multilayer perceptron to predict the final class label. We utilized a straightforward multilayer perceptron (MLP) architecture, comprising two hidden layers with 256 and 128 neurons, respectively, each activated by ReLU functions.
\subsection{Hyper-parameters}
We utilized cross-entropy loss and Adam optimizer with decoupled weight decay \cite{adamw}, setting the initial learning rate at $10^{-2}$ and applying weight decay regularization of $10^{-3}$. Additionally, a scheduling tool was incorporated for the reduction of learning rates on a plateau, and different batch sizes (64, 128, and 256) were tested. ResNets are trained using $100$ epochs, while the MLPs are trained on $500$ epochs.
\subsection{Metrics for Performance Evaluation}
In our evaluation, we utilized several metrics to comprehensively assess the performance of our model: Accuracy, Weighted F1 Score, F1 Score, and Area Under the Curve (AUC)\cite{fawcett2006introduction}. Accuracy provides a straightforward measure of the proportion of correctly classified instances. However, it can be misleading in the presence of class imbalance. To address this, we included the Weighted F1 Score, which accounts for both precision and recall across different classes, giving more importance to classes with a higher number of instances. The standard F1 Score was also used to evaluate the balance between precision and recall for the minority class. Finally, AUC was chosen to measure the ability of the architecture to distinguish between classes, offering a robust evaluation metric. AUC resulted in $99.$
\section{Results and Discussions}
\begin{table*}[h]
    \centering\small
    \begin{tabular}{c|ccccc}
        & Accuracy &Weighted F1-score & F1-score  &  AUC \\
        \hline
        AVES-bio \cite{hagiwara2023aves}&   87.90\% & - & - & -  \\
        ResNet \cite{murphy2022residual}&   - & - & 86.70\% & 92.80\% \\
        Transfer Learning \cite{ghani2023global}  & 83.00\% & - & - & 98.00\%\\
        BEANS \cite{hagiwara2023beans} &   87.00\% & - & - & -  \\
        \hline
       \textbf{WhaleNet} w/ max merge  &   96.67\% & 96.62\% & 93.04\% & 99.75\% \\
        \textbf{WhaleNet} w/ hard merge    & \textbf{97.60\%} & \textbf{97.61\%}  & \textbf{93.81\%}&\textbf{99.82\%}\\
        \textbf{WhaleNet} w/ MLP merge    & {96.35\%} & {96.37\%}  & {91.89\%}&{99.71\%}\\
    \end{tabular}
    \caption{\textit{Main Results}. Considering the classification task on the entire WMMD dataset \cite{sayigh2016watkins}, we compare state-of-the-art benchmarks with our proposal (last two rows). The optimal batch size was found to be $128$, and the WST configuration was set to $(J,Q) = (6,16)$. We report, when available for existing results, standard performance metrics such as accuracy, F1 score, and AUC score, calculated in the test set at the end of training. Since the number of elements per class varies significantly, we also report the weighted F1 score. The top performance for each metric is emphasized. Disclaimer: \cite{hagiwara2023beans}, \cite{murphy2022residual}, \cite{ghani2023global} use only the "best of" subset of the full dataset, cfr. \cite{sayigh2016watkins}, while we use the full dataset. In \cite{hagiwara2023aves} is not specified. }
    \label{tab:results}
\end{table*}
Table \ref{tab:results} shows the quantitative performance metrics of the WhaleNet architecture employing three distinct merging strategies, in comparison to extant benchmarks. Given the pronounced class imbalance within the dataset, a weighted F1-score was computed, accounting for the number of elements per class. Evidently, our pre-processing pipeline and WhaleNet surpass state-of-the-art models by almost $9-10\%$, achieving accuracies of $96.67\%$, $97.60\%$, and $96.35\%$ with \textit{max merge}, \textit{hard merge}, and \textit{MLP merge}, respectively, thereby exceeding the symbolic threshold of $90\%$. While this may appear as a marginal absolute improvement, the reduction in misclassification rates from approximately $12\%$ in benchmark models to less than half in our proposal signifies a substantial advancement in addressing the classification task for the dataset under study. Furthermore, the results of additional experiments, detailed in Appendix A, demonstrate that WhaleNet, even with varying hyper-parameter configurations, such as different values of $J$ and $Q$ for the WST, and different batch sizes, consistently outperforms the current state-of-the-art. 

An examination of the results presented in Appendix A (Table \ref{tab:wst6,16} and \ref{tab:wst7,10}) elucidates the underlying principles of the WhaleNet architecture. Notably, it is imperative to highlight that each ResNet block, whether trained on WST or Mel representation, surpasses state-of-the-art performance by an average margin of $7\%$ in accuracy. Nevertheless, WST and Mel spectrograms accentuate disparate features, and, importantly, the second-order WST exhibits the capability to achieve more pronounced distinctions in the data. In order to increase classification outcomes and thereby furnish a more robust architecture suitable for practical and real-world intelligent systems, the application of ensemble learning yields exceptional results. This resulted in an enhancement in accuracy exceeding $2\%$, culminating in an overall accuracy of $98\%$. Furthermore, an examination of the final results presented in Table \ref{tab:results} reveals that additional metrics, such as the F1-score and AUC, also outperformed state-of-the-art models for whale vocalization. Notably, WhaleNet, when employing all three final merging layers, consistently achieved an accuracy greater than $99.70\%$, with the F1-score surpassing state-of-the-art benchmarks by more than $6\%$.

\section{Conclusions}
In this study, we focus on the Watkins Marine Mammal Sound Database (WMMD), a comprehensive and labeled dataset of marine mammal vocalizations. Due to its pronounced imbalance and heterogeneity in terms of signal length, data preparation, and classification tasks posed considerable challenges. In this paper, we initially introduced a clear and straightforward data preparation pipeline, employing a time-frequency analysis based on Mel spectrograms, a standard approach, in contrast to an alternative method based on Wavelet Scattering Transform (WST). Subsequently, we introduce WhaleNet architecture to specifically address a classification task on the entire dataset, a deep learning model that uses residual layers and ensembles the different information provided by WST and Mel spectrogram. Our model surpassed state-of-the-art accuracy results by almost $10\%$, achieving accuracy values of $97.60\%$ of correct predictions. The accuracy reached is notable, especially considering the heterogeneity of the dataset, both in signal length and class distribution. In addition, existing work usually focused only on subsets of the full dataset. Given this performance, we conclude that the precision of our method can be of fundamental interest for bioacoustics, bridging the gap between the data science and biology communities. Furthermore, the analyzed dataset itself serves as a crucial case study for machine learning applications to natural datasets. Based on the results presented, future directions could involve a more thorough investigation of optimal parameter pairs $(J,Q)$ and other hyperparameters of the model. It would be possible to implement a majority voting routine that simultaneously considers two parallel networks trained on WST and Mel spectrogram. However, we believe that any further improvement in accuracy would necessitate better class balancing, potentially through additional measurements or data augmentation for the less-represented species in the dataset. With these adjustments, a near-perfect classification could be within reach.

\section*{Acknowledgments}
D.C. and A.L. worked under the auspices of Italian National Group of Mathematical Physics (GNFM) of INdAM. A.L. is part of the project PNRR-NGEU which has received funding from the MUR – DM 117/2023, and was supported in part through the Politecnico di Torino IT High Performance Computing resources, services, and staff expertise. D.C. was supported in part through the NYU IT High Performance Computing resources, services, and staff expertise.



\bibliography{example_paper}
\bibliographystyle{icml2024}

\newpage
\appendix
\onecolumn
\section{Additional Experimental Results.}
\label{app:A}
\begin{figure}[h]
    \centering
    \includegraphics[width = .4\linewidth]{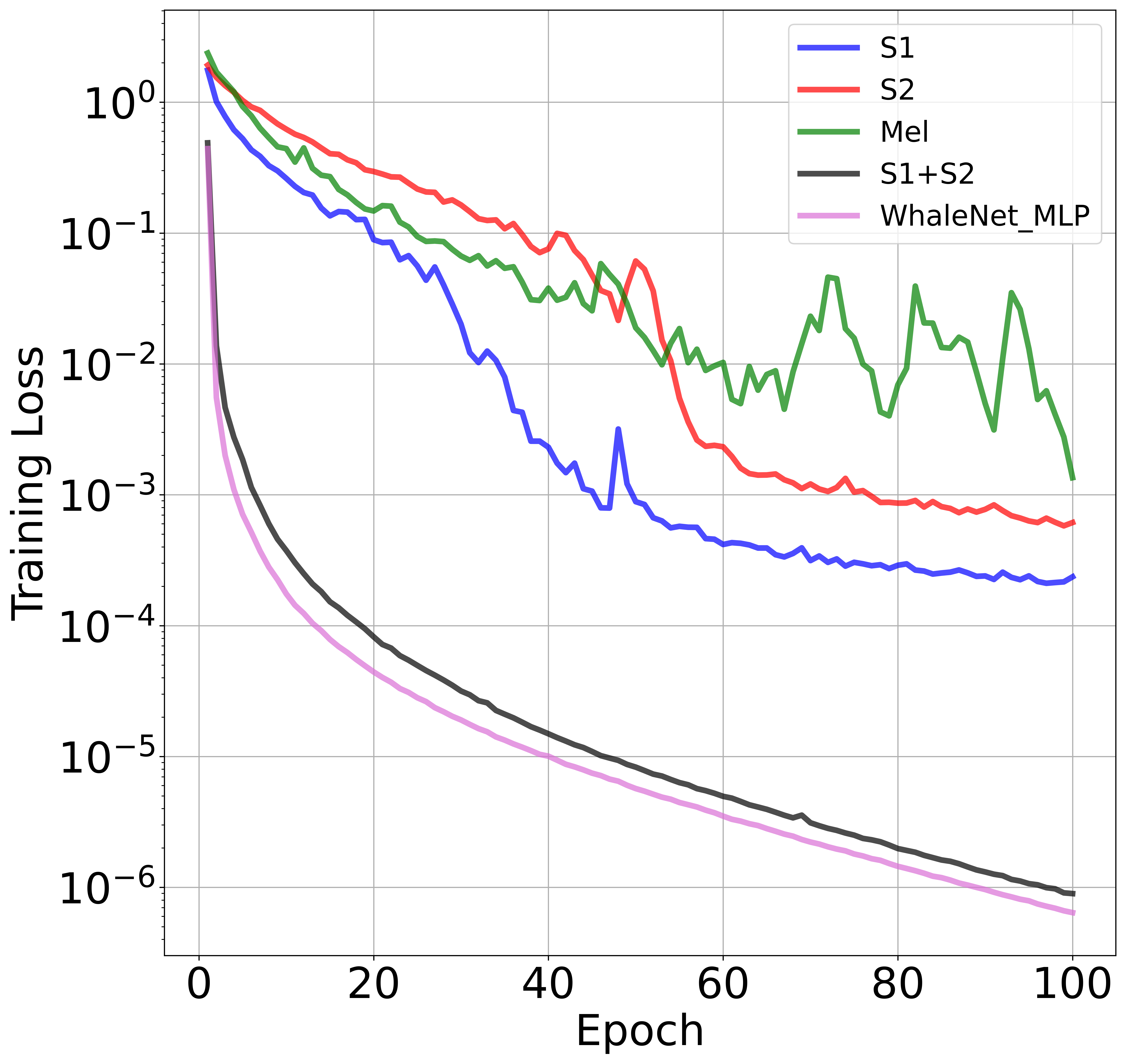}
    \includegraphics[width = .4\linewidth]{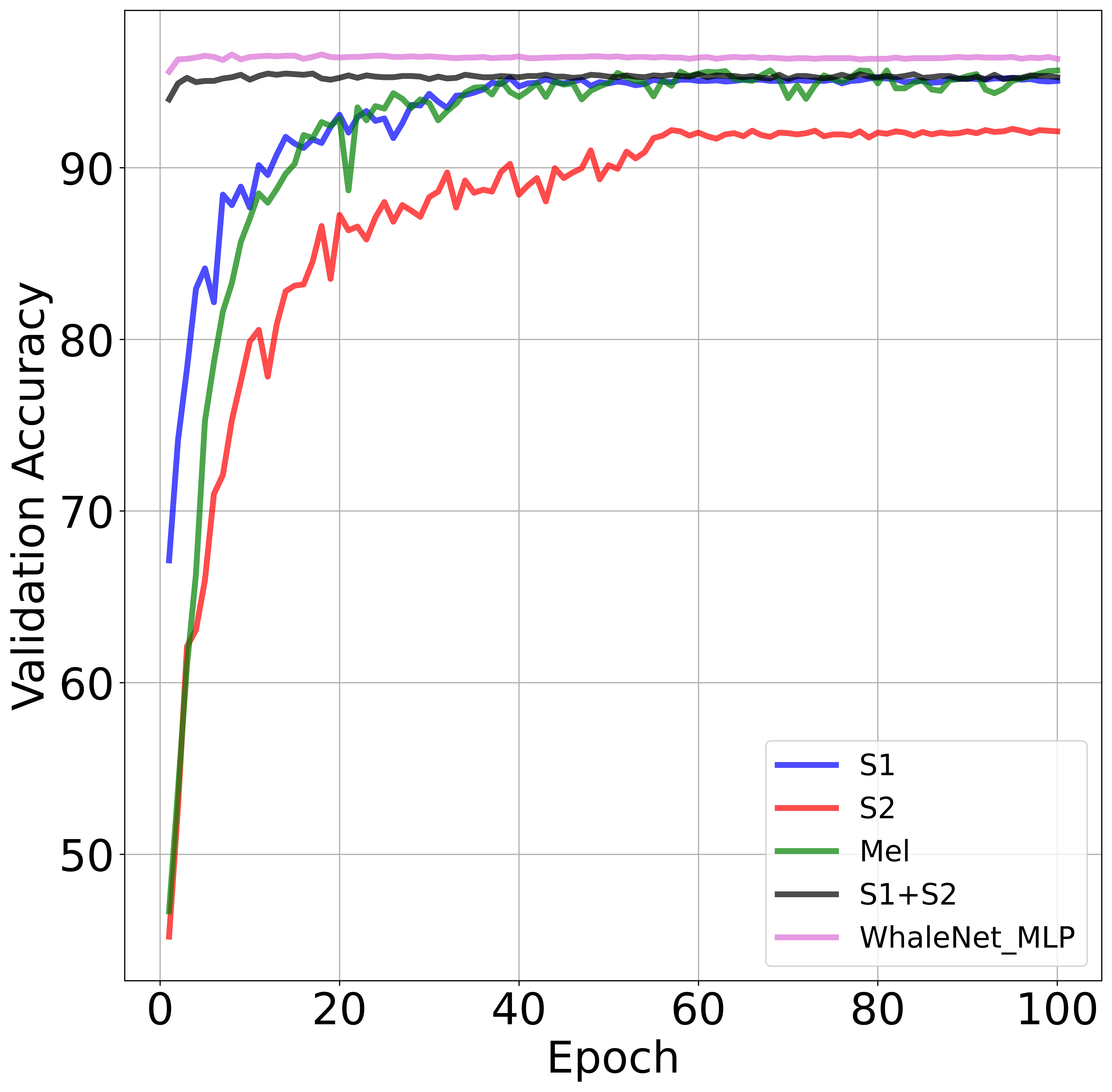}
    \caption{Some experiments on the best combination of hyper-parameters, i.e. $(J,Q) = (6,16)$ and batch size equal to $128$. \textit{Left} The training loss (in logarithmic scale) of WhaleNet branches over the initial 100 epochs tells that the MLPs employed to integrate WST order 1, WST order 2, and Mel, respectively, facilitated a more seamless convergence on the training set, resulting in a reduced loss. \\
  \textit{Right} The validation accuracy of WhaleNet branches over the initial 100 epochs tells the MLPs used for the integration of WST order 1, WST order 2, and Mel-frequency cepstral coefficients  have collectively contributed to an enhancement in overall accuracy. 
    }
    \label{test_128}
\end{figure}

In this section, we present supplementary experimental results. Figure \ref{test_128} illustrates the test loss and validation accuracy per epoch for a batch size of $128$ across each trainable branch of WhaleNet. Additionally, Tables \ref{tab:wst6,16} and \ref{tab:wst7,10} enumerate the accuracy scores achieved by the model architecture under various hyperparameters, specifically batch sizes, merging algorithms, and distinct values of the $J$ and $Q$ parameters of the WST. Further plots and results are available on the github repository \href{https://github.com/alelicciardi99/whalenet/tree/main}{\texttt{whalenet\_vocalization\_classification}}.

\begin{table*}[h]
    \centering\small
    \begin{tabular}{c|c|c c c | c c c}
    
    Batch Size & Mel & S1 & S2 & S1+S2 & Max Merge & Hard Merge & MLP Merge\\
    \hline
    64 & 95.67\% & 95.06\% & 92.12\% & 95.23\% & 96.41\% & 97.24\%&96.27\%\\
    128 & 95.91\% & 95.09\% & 92.66\% & 95.63\% & 96.56\% & 97.60\%&96.35\%\\
    256 & 95.23\% & 95.70\% & 91.94\% & 95.63\% & 96.67\% & 97.28\%&96.31\%\\
        
    \end{tabular}
    \vspace{2ex}
\caption{Performance metrics in terms of accuracy on the validation dataset post-training. Various combinations of hyperparameters are presented, specifically the batch size and the merging methodologies, namely max merge, hard merge, and MLP merge. The WST was calculated with $J,Q$ values set to $6$ and $16$.}
\label{tab:wst6,16}
\end{table*}

\begin{table*}[h]
    \centering\small
    \begin{tabular}{c|c|c c c | c c c}
    
    Batch Size & Mel & S1 & S2 & S1+S2 & Max Merge & Hard Merge & MLP Merge\\
    \hline
    64 & 95.91\% & 94.56\% & 91.48\% & 94.77\% & 96.45\% & 97.17\%&95.34\%\\
    128 & 95.67\% & 94.59\% & 91.51\% & 94.20\% & 95.77\% & 96.77\%&95.45\%\\ 
    256 & 95.23\% & 95.02\% & 91.62\% & 95.20\% & 95.41\% & 96.78\%&96.06\%\\
        
    \end{tabular}
    \vspace{2ex}
\caption{Performance metrics in terms of accuracy on the validation dataset post-training. Various combinations of hyperparameters are presented, specifically the batch size and the merging methodologies, namely max merge, hard merge, and MLP merge. The WST was calculated with $J,Q$ values set to $7$ and $10$.}
\label{tab:wst7,10}
\end{table*}

\end{document}